\documentclass[10pt,twocolumn]{IEEEtran}
\usepackage{amsmath,amssymb,epsfig,psfrag,cite,subfigure}
\include{macros}
\usepackage{graphicx}

\usepackage{algorithm,mathtools}
\usepackage{epsfig,psfrag}
\usepackage{subfigure}
\usepackage{color}
\usepackage{url}

\newcommand{\mtH}{{\mathcal{H}}}
\newcommand{\mtN}{{\mathcal{N}}}
\newcommand{\mtS}{{\mathcal{S}}}

\newcommand{\rmP}{{\rm{P}}}

\begin{document}

\title{On the Optimality of Likelihood Ratio Test for Prospect Theory Based Binary Hypothesis Testing}

\author{Sinan Gezici,\thanks{S. Gezici is with the Dept. of Electrical and Electronics Eng.,
Bilkent University, Bilkent, Ankara 06800, Turkey, Tel: +90-312-290-3139, E-mail:
gezici@ee.bilkent.edu.tr. P. K. Varshney is with the Dept. of Electrical Engineering
and Computer Science, Syracuse University, Syracuse, NY 13244, USA, E-mail: varshney@syr.edu. P. K. Varshney's work was supported by AFOSR Grant FA9550-17-1-0313 under the DDDAS program.} \emph{Senior Member, IEEE}, and Pramod K. Varshney, \emph{Life Fellow, IEEE}\vspace{-0.5cm}}

\maketitle

\begin{abstract}
  In this letter, the optimality of the likelihood ratio test (LRT) is investigated for binary hypothesis testing problems in the presence of a behavioral decision-maker. By utilizing prospect theory, a behavioral decision-maker is modeled to cognitively distort probabilities and costs based on some weight and value functions, respectively. It is proved that the LRT may or may not be an optimal decision rule for prospect theory based binary hypothesis testing and conditions are derived to specify different scenarios. In addition, it is shown that when the LRT is an optimal decision rule, it corresponds to a randomized decision rule in some cases; i.e., nonrandomized LRTs may not be optimal. This is unlike Bayesian binary hypothesis testing in which the optimal decision rule can always be expressed in the form of a nonrandomized LRT. Finally, it is proved that the optimal decision rule for prospect theory based binary hypothesis testing can always be represented by a decision rule that randomizes at most two LRTs. Two examples are presented to corroborate the theoretical results.

\end{abstract}

\vspace{-0.4cm}

\section{Introduction}\label{sec:Intro}


In hypothesis testing problems, a decision-maker aims to design an optimal decision rule according to a certain approach such as the Bayesian, minimax, or Neyman-Pearson (NP) \cite{Poor,Levy}. In the presence of prior information, the Bayesian approach is commonly employed, where the decision-maker wishes to minimize the average cost of making decisions, i.e., the \emph{Bayes risk}. The calculation of the Bayes risk requires the knowledge of costs of possible decisions and probabilities of possible events. However, in case of a human decision-maker, such knowledge may not be perfectly available due to both limited availability of information and/or complex human behaviors such as emotions, loss-aversion, and endowment effect (see \cite{PTbased2016} and references therein). The behavior of a human decision-maker is effectively modeled via \emph{prospect theory}, which utilizes weight and value functions to capture the impact of human behavior on probabilities and costs \cite{PTmain}. In prospect theory based hypothesis testing, the aim of a human decision-maker (also known as behavioral decision-maker) becomes the minimization of the \emph{behavioral risk}, which generalizes the Bayes risk by transforming probabilities and costs according to the behavioral parameters of the decision-maker.

Recently, optimal decision rules are investigated in \cite{PTbased2016} for binary hypothesis testing problems when decision-makers are modeled via prospect theory. Two special types of behavioral decision-makers, namely optimists and pessimists, are considered, and a known (concave) relation is assumed between the false alarm and detection probabilities of a decision-maker. It is shown that the optimal decision rule can achieve different false alarm and detection probabilities than those attained by the Bayes decision rule in the presence of a behavioral decision-maker. In a related work, a game theoretic problem is formulated for strategic communications between a human transmitter and a human receiver, which are modeled via prospect theory \cite{Basar_StraCom}. It is found that behavioral decision-makers employ the same equilibrium strategies as those for non-behavioral (unbiased) decision-makers in the Stackelberg sense.

The aim of this letter is to derive optimal decision rules for generic behavioral decision-makers in binary hypothesis testing problems. To that aim, the optimality of the likelihood ratio test (LRT), which is known to be the optimal decision rule in the Bayesian framework, is investigated for prospect theory based binary hypothesis testing. It is proved that the LRT may or may not be an optimal decision rule for behavioral decision-makers, and conditions are provided to specify various scenarios. In addition, it is shown that when the LRT is an optimal decision rule, it corresponds to a randomized LRT in some cases. This is different from the Bayesian approach in which the optimal decision rule can always be stated as a nonrandomized LRT. Finally, the generic solution of the prospect theory based binary hypothesis testing problem is obtained; namely, it is proved that the optimal solution can always be represented by randomization of at most two LRTs. Two classical examples are used to support the theoretical results.

\vspace{-0.2cm}

\section{Problem Formulation and Theoretical Results}\label{sec:SysModel}

Consider a binary hypothesis testing problem in the presence of a behavioral decision-maker \cite{PTbased2016}. The hypotheses are denoted by $\mtH_0$ and $\mtH_1$, and the prior probabilities are given by $\pi_0=\rmP(\mtH_0)$ and $\pi_1=\rmP(\mtH_1)$. The observation of the decision-maker is $r\in\Gamma$, where $\Gamma$ represents the observation space. Observation $r$ has conditional distributions {$p_0(r)$ and $p_1(r)$} under $\mtH_0$ and $\mtH_1$, respectively. The  behavioral decision-maker employs a decision rule $\phi(r)$ to determine the true hypothesis, where $\phi(r)$ corresponds to the probability of selecting $\mtH_1$; that is, $\phi:\Gamma\rightarrow [0,1]$.

As in \cite{PTbased2016}, the rationality of the decision-maker is modeled via prospect theory \cite{PTmain}, {\cite{PTadv}} in this work. {In prospect theory, loss aversion, risk-seeking and risk-aversion behaviors of humans are characterized, where a} behavioral decision-maker cognitively distorts the probabilities and costs based on some known weight function $w(\cdot)$ and value function $v(\cdot)$, respectively \cite{PTmain}, {\cite{PTadv,PTyears}}. Then, the classical \emph{Bayes risk} for binary hypothesis testing becomes the following \emph{behavioral risk} for prospect theory based binary hypothesis testing \cite{PTbased2016}:
\begin{gather}\label{eq:BehRisk}
f(\phi)=\sum_{i=0}^{1}\sum_{j=0}^{1}w\big(\rmP(\mtH_i{\rm{~selected}}\,\&\,\mtH_j{\rm{~true}})\big)v(c_{ij})
\end{gather}
where $c_{ij}$ is the cost of deciding in favor of $\mtH_i$ when the true hypothesis is $\mtH_j$ \cite{Poor}. It is noted that the behavioral risk in \eqref{eq:BehRisk} reduces to the Bayes risk for $w(p)=p$ and $v(c)=c$.

The aim of the decision-maker is to find the optimal decision rule $\phi^*$ that minimizes the behavioral risk in \eqref{eq:BehRisk}; that is; to solve the following optimization problem:
\vspace{-0.15cm}\begin{gather}\label{eq:optProb}
\phi^*(r)=\arg\underset{\phi}\min\,f(\phi)
\end{gather}

\vspace{-0.2cm}

\noindent To that aim, the following relation can be utilized first,  $\rmP(\mtH_i{\rm{~selected}}\,\&\,\mtH_j{\rm{~true}})=\pi_j\,\rmP(\mtH_i{\rm{~selected}}\,|\,\mtH_j{\rm{~true}})$, and \eqref{eq:BehRisk} can be written as
\vspace{-0.2cm}\begin{align}\label{eq:BehRisk2}
f(\phi)&=g(x)+h(y)\\\label{eq:g}
g(x)&=w(\pi_0(1-x))v(c_{00})+w(\pi_0x)v(c_{10})\\\label{eq:h}
h(y)&=w(\pi_1(1-y))v(c_{01})+w(\pi_1y)v(c_{11})
\end{align}

\vspace{-0.15cm}

\noindent where $x=\int_{\Gamma}\phi(r){p_0(r)}dr$ and $y=\int_{\Gamma}\phi(r){p_1(r)}dr$ are the false alarm and detection probabilities, respectively \cite{PTbased2016}. Then, the following proposition states the (non-)optimality of the LRT under various conditions.

\textit{\textbf{Proposition 1:} Suppose that the weight function $w(\cdot)$ is monotone increasing.}

\textit{{Case~(a):} If $v(c_{10})v(c_{00})<0$ or $v(c_{11})v(c_{01})<0$, then the LRT is a solution of \eqref{eq:optProb}.}

\textit{{Case~(b):} If $v(c_{10})v(c_{00})\geq0$ and $v(c_{11})v(c_{01})\geq0$, then the LRT may or may not be a solution of \eqref{eq:optProb}.}

\indent\indent{\textit{Proof:}} {\underline{Case~(a):}} Consider the {scenario} in which $v(c_{10})>0$ and $v(c_{00})<0$. Let $\phi'$ denote an arbitrary decision rule, which achieves false-alarm probability $x'$ and detection probability $y'$. Then, define $\phi^*_1$ as an LRT given by
\vspace{-0.2cm}\begin{gather}\label{eq:DecRuleLRT1}
\phi^*_1(r)=\begin{cases}
0\,,&{\rm{if}}~{p_1(r)< \eta\, p_0(r)}\\
\gamma\,,&{\rm{if}}~{p_1(r)=\eta\, p_0(r)}\\
1\,,&{\rm{if}}~{p_1(r)> \eta\, p_0(r)}
\end{cases}
\end{gather}

\vspace{-0.15cm}

\noindent where $\eta\geq0$ and $\gamma\in[0,1]$ are chosen such that the detection probability of $\phi^*_1$ is equal to $y'$. Then, similar to the proof of the Neyman-Pearson lemma \cite[p.~24]{Poor}, the following relation can be derived based on \eqref{eq:DecRuleLRT1}:
\vspace{-0.2cm}\begin{align}\label{eq:intIneq1}
\int_{\Gamma}\big({p_1(r)-\eta\,p_0(r)}\big)\left(\phi^*_1(r)-\phi'(r)\right)dr\geq 0\,.
\end{align}

\vspace{-0.15cm}

\noindent From \eqref{eq:intIneq1}, $\eta(x^*-x')\leq y^*-y'$ is obtained, where $x^*$ and $y^*$ represent the false-alarm and detection probabilities of $\phi^*_1$, respectively. Since the detection probability of $\phi^*_1$ is set to $y'$ and $\eta\geq0$, it is concluded that $x^*\leq x'$. Hence, for any decision rule $\phi'$, the LRT in \eqref{eq:DecRuleLRT1} achieves an equal or lower false alarm probability for the same level of detection probability. This means that the use of the LRT can reduce $g(x)$ in \eqref{eq:g} (as $v(c_{10})>0$, $v(c_{00})<0$, and $w(\cdot)$ is monotone increasing) without changing the value of $h(y)$ in \eqref{eq:h}. Therefore, it is deduced that no other test can achieve a lower behavioral risk (see \eqref{eq:BehRisk2}) than the LRT in \eqref{eq:DecRuleLRT1}.\footnote{The existence of \eqref{eq:DecRuleLRT1} can be proved similarly to the proof of the Neyman-Pearson lemma \cite[pp.~24-25]{Poor}.}

Now suppose that $v(c_{10})<0$ and $v(c_{00})>0$, and again let $\phi'$ denote an arbitrary decision rule, which achieves false-alarm probability $x'$ and detection probability $y'$. In this {scenario}, define $\phi^*_2$ as an LRT that is stated as
\vspace{-0.2cm}\begin{gather}\label{eq:DecRuleLRT2}
\phi^*_2(r)=\begin{cases}
0\,,&{\rm{if}}~{p_1(r)>\beta\,p_0(r)}\\
\kappa\,,&{\rm{if}}~{p_1(r)= \beta\, p_0(r)}\\
1\,,&{\rm{if}}~{p_1(r)< \beta\, p_0(r)}
\end{cases}
\end{gather}

\vspace{-0.15cm}

\noindent where $\beta\geq0$ and $\kappa\in[0,1]$ are chosen such that the detection probability of $\phi^*_2$ is equal to $y'$. Then, it can be shown that
\vspace{-0.2cm}\begin{align}\label{eq:intIneq2}
\int_{\Gamma}\big(\beta\,{p_0(r)-p_1(r)}\big)\left(\phi^*_2(r)-\phi'(r)\right)dr\geq 0
\end{align}

\vspace{-0.2cm}

\noindent which leads to $\beta(x^*-x')\geq y^*-y'$. Therefore, $x^*\geq x'$ is obtained as $y^*=y'$ and $\beta\geq0$. Hence, for any decision rule $\phi'$, the LRT in \eqref{eq:DecRuleLRT2} achieves an equal or higher false alarm probability for the same level of detection probability. This implies that no other test can achieve a lower behavioral risk than the LRT in \eqref{eq:DecRuleLRT2} since $v(c_{10})<0$, $v(c_{00})>0$, and $w(\cdot)$ is monotone increasing (see \eqref{eq:BehRisk2}--\eqref{eq:h}).

For $v(c_{11})<0$ and $v(c_{01})>0$, similar arguments can be employed to show that for any arbitrary decision rule $\phi'$ with false-alarm probability $x'$ and detection probability $y'$, an LRT in the form of \eqref{eq:DecRuleLRT1} can be designed to achieve the same false-alarm probability but an equal or higher detection probability. Since $v(c_{11})<0$ and $v(c_{01})>0$ in this {scenario,} $h(y)$ can be reduced without affecting $g(x)$. Therefore, no other test can achieve a lower behavioral risk than the LRT.

For $v(c_{11})>0$ and $v(c_{01})<0$, it can be shown that for an arbitrary decision rule $\phi'$ with false-alarm probability $x'$ and detection probability $y'$, an LRT in the form of \eqref{eq:DecRuleLRT2} can be designed to achieve the same false-alarm probability but an equal or lower detection probability. Since $v(c_{11})>0$ and $v(c_{01})<0$, $h(y)$ can be reduced without affecting $g(x)$. Hence, no other test can achieve a lower behavioral risk than the LRT.

{\underline{Case~(b):}} It suffices to provide examples in which the LRT is and is not a solution of \eqref{eq:optProb}. First, consider a scenario in which the weight function is given by $w(p)=p$ for $p\in[0,1]$. Then, the behavioral risk becomes the classical Bayes risk (by defining $v(c_{ij})$'s as new costs). Hence, the optimal decision rule is given by the LRT \cite[pp.~6-7]{Poor}, which is in the form of \eqref{eq:DecRuleLRT1} or \eqref{eq:DecRuleLRT2}. Next, for an example in which the LRT is not a solution of \eqref{eq:optProb}, please see Section~\ref{sec:Ex1}.\hfill$\blacksquare$

Proposition 1 reveals that when the probabilities are distorted by a behavioral decision-maker, the LRT may lose its optimality property for binary hypothesis testing {when both $v(c_{10})v(c_{00})\geq0$ and $v(c_{11})v(c_{01})\geq0$ are satisfied. It is also noted that having at least one of $v(c_{10})v(c_{00})<0$ or $v(c_{11})v(c_{01})<0$ is a sufficient condition for the optimality of the LRT.}

The signs of the $v(c_{ij})$ terms are determined depending on whether the behavioral decision-maker perceives the cost of choices as detrimental or profitable. In particular, if selecting $\mtH_i$ when $\mtH_j$ is true is perceived as detrimental (profitable), then $v(c_{ij})\geq0$ ($v(c_{ij})\leq0$) \cite{PTbased2016}. Therefore, perceptions of a decision-maker can affect the optimality of the LRT in prospect theory based binary hypothesis testing. (For example, in \emph{strategic} information transmission, various cost perceptions can be observed depending on utilities of decision-makers \cite{SerkanHT}.)

\textit{\textbf{{Remark~1:}}
In most experimental studies, the weight function is observed to behave in a monotone increasing manner \cite{Weight2}, \cite{Weight1}; hence, the assumption in the proposition holds commonly.}

It is well-known that the optimal decision rule can always be expressed in the form of a nonrandomized LRT for Bayesian hypothesis testing \cite{Poor}. In other words, according to the Bayesian criterion (which aims to minimize \eqref{eq:BehRisk} for $w(p)=p$ and $v(c)=c$), the optimal decision rule is to compare the likelihood ratio against a threshold and to choose $\mtH_0$ or $\mtH_1$ arbitrarily whenever the likelihood ratio is equal to the threshold (i.e., randomization is not necessary). However, for prospect theory based hypothesis testing, randomization can be required to obtain the optimal solution (i.e., the solution of \eqref{eq:optProb}) in some scenarios. This is stated in the following.

\textit{\textbf{{Remark~2:}} Suppose that the solution of \eqref{eq:optProb} is in the form of an LRT; that is, \eqref{eq:DecRuleLRT1} or \eqref{eq:DecRuleLRT2}. Then, in some cases, the optimal decision rule may need to be randomized with the randomization constant being in the open interval $(0,1)$.}

{To justify Remark~2}, consider $w(p)=p$ and $v(c)=c$ in \eqref{eq:BehRisk}; that is, the Bayesian framework. Then, a nonrandomized LRT (i.e., \eqref{eq:DecRuleLRT1} with $\gamma\in\{0,1\}$ or \eqref{eq:DecRuleLRT2} with $\kappa\in\{0,1\}$) is always an optimal solution of \eqref{eq:optProb} \cite{Poor}. Next, consider the example in Section~\ref{sec:Ex2}, where the optimal solution is in the form of \eqref{eq:DecRuleLRT1} with $\gamma\in(0,1)$ (see \eqref{eq:OptRandSol}); hence, no nonrandomized decision rules can be a solution of \eqref{eq:optProb} in certain scenarios.

Finally, the optimal decision rule is specified for prospect theory based binary hypothesis testing in the general case. To that aim, the problem in \eqref{eq:optProb} is stated, based on \eqref{eq:BehRisk2}-\eqref{eq:h}, as
\begin{gather}\label{eq:optProb2}
(x^*,y^*)=\underset{(x,y)\in\mtS}{\rm{arg\,min}}~g(x)+h(y)
\end{gather}
where $\mtS$ denotes the set of achievable false alarm and detection probabilities for the given problem, and $x^*$ and $y^*$ represent, respectively, the false alarm and detection probabilities attained by the optimal decision rule in \eqref{eq:optProb}. Once the problem in \eqref{eq:optProb2} is solved, any decision rule with false alarm probability $x^*$ and detection probability $y^*$ becomes an optimal solution. The following proposition states that the optimal solution can always be represented by a decision rule that performs randomization between at most two LRTs.

\textit{\textbf{{Proposition 2:}} The solution of \eqref{eq:optProb} can be expressed as a randomized decision rule which employs $\phi^*_1$ in \eqref{eq:DecRuleLRT1} with probability $(y^*-y^*_2)/(y^*_1-y^*_2)$ and $\phi^*_2$ in \eqref{eq:DecRuleLRT2} with probability $(y^*_1-y^*)/(y^*_1-y^*_2)$, where $y^*_1$ ($y^*_2$) is the detection probability of $\phi^*_1$ ($\phi^*_2$) when its false alarm probability is set to $x^*$, and $x^*$ and $y^*$ are given by \eqref{eq:optProb2}.}

\indent\indent{\textit{Proof:}} Consider the optimization problem in \eqref{eq:optProb2}, the solution of which is denoted by $(x^*,y^*)$. It is known that $\mtS$ is a convex set in $[0,1]\times[0,1]$ \cite[p.~33]{Levy}.\footnote{Therefore, \eqref{eq:optProb2} becomes a convex optimization problem if $g(x)$ is a convex function of $x$ and $h(y)$ is a convex function of $y$.} Since $\phi^*_1$ in \eqref{eq:DecRuleLRT1} attains the maximum detection probability for any given false alarm probability (as discussed in the proof of Proposition~1), the upper boundary of $\mtS$ is achieved by $\phi^*_1$. Similarly, the lower boundary of $\mtS$ is achieved by $\phi^*_2$ in \eqref{eq:DecRuleLRT2} as it provides the minimum detection probability for any given false alarm probability. Design the parameters of $\phi^*_1$ in \eqref{eq:DecRuleLRT1} and $\phi^*_2$ in \eqref{eq:DecRuleLRT2} such that their false alarm probabilities become equal to $x^*$, and let $y^*_1$ and $y^*_2$ represent their corresponding detection probabilities. Due to the previous arguments, $y^*_1\geq y^*\geq y^*_2$ holds. Choose $\nu=(y^*-y^*_2)/(y^*_1-y^*_2)$ and randomize $\phi^*_1$ and $\phi^*_2$ with probabilities $\nu$ and $1-\nu$, respectively. Then, the resulting randomized decision rule attains a detection probability of $y^*$ and a false alarm probability of $x^*$. Therefore, it becomes the solution of \eqref{eq:optProb2}; hence, the optimal decision rule according to \eqref{eq:optProb}.\hfill$\blacksquare$

{Proposition 2} implies that the optimal decision rule for prospect theory based binary hypothesis testing can be expressed in terms of the LRT in \eqref{eq:DecRuleLRT1} (if $y^*=y^*_1$), the LRT in \eqref{eq:DecRuleLRT2} (if $y^*=y^*_2$), or their randomization (if $y^*\in(y^*_1,y^*_2)$). It should be noted that the randomization of two LRTs is not in the form of an LRT in general. Hence, the LRT may or may not be an optimal decision rule, as stated in Proposition 1.

{By deriving the optimal decision rules for prospect theory based hypothesis testing, we provide theoretical performance bounds for behavioral (human) decision-makers. As humans may not implement these optimal rules exactly in practice, we can evaluate how close to optimal they perform.}

{\textit{\textbf{Remark~3:} Randomized decision rules generalize deterministic decision rules and can outperform them in certain scenarios (e.g., \cite[pp.~27--29]{Poor}, \cite{MUdetRand,detRandBD}).}}

\vspace{-0.18cm}

\section{Examples {and Conclusions}}\label{sec:Examples}

In this section, two classical problems in binary hypothesis testing are investigated from a prospect theory based perspective. {For the weight function, the following commonly used model in prospect theory is employed \cite{Weight2,Weight1,PTadv}:
\vspace{-0.15cm}\begin{gather}\label{eq:weight}
w(p)=\frac{p^{\alpha}}{(p^{\alpha}+(1-p)^{\alpha})^{1/\alpha}}\,,~p\in[0,1]~{\textrm{and}}~\alpha>0
\end{gather}
where $\alpha$ is a probability distortion parameter of the decision-maker. The model in \eqref{eq:weight} is supported via various experiments and it can capture risk-seeking and risk-aversion attitudes of human decision-makers in different scenarios \cite{Weight2,Weight1,PTadv}.}

\vspace{-0.1cm}

\subsection{Example 1: Location Testing with Gaussian Error}\label{sec:Ex1}

Suppose observation $r$ is a scalar random variable distributed as $\mtN(\mu_i,\sigma^2)$ under hypothesis $\mtH_i$ for $i\in\{0,1\}$, where $\mtN(\mu_i,\sigma^2)$ denotes a Gaussian random variable with mean $\mu_i$ and variance $\sigma^2$. For this hypothesis testing problem, the LRTs in \eqref{eq:DecRuleLRT1} and \eqref{eq:DecRuleLRT2} can be stated as follows:
\vspace{-0.15cm}\begin{gather}\label{eq:LRTex1}
\phi^*_1(r)=\begin{cases}
0\,,&\hspace{-0.1cm}{\rm{if}}~r<\tau\\
1\,,&\hspace{-0.1cm}{\rm{if}}~r\geq \tau
\end{cases}
,~
\phi^*_2(r)=\begin{cases}
1\,,&\hspace{-0.1cm}{\rm{if}}~r<\tilde{\tau}\\
0\,,&\hspace{-0.1cm}{\rm{if}}~r\geq\tilde{\tau}
\end{cases}.
\end{gather}

\vspace{-0.15cm}

\noindent The corresponding false alarm and detection probabilities can be obtained, respectively, as $x=Q\big(\frac{\tau-\mu_0}{\sigma}\big)$ and $y=Q\big(\frac{\tau-\mu_1}{\sigma}\big)$ for the first LRT in \eqref{eq:LRTex1} and as $x=Q\big(\frac{\mu_0-\tilde{\tau}}{\sigma}\big)$ and $y=Q\big(\frac{\mu_1-\tilde{\tau}}{\sigma}\big)$ for the second LRT in \eqref{eq:LRTex1}.

It is well-known that the LRT is the optimal decision rule according to the Bayesian criterion \cite[pp.~11-12]{Poor}. To show that it may not always be optimal in the prospect theory based framework, consider the optimal decision rule that is specified, based on {Proposition 2}, as follows:
\vspace{-0.15cm}\begin{gather}\label{eq:optOverall}
\phi^*(r)=\nu\,\phi^*_1(r)+(1-\nu)\phi^*_2(r)
\end{gather}
where $\nu=(y^*-y^*_2)/(y^*_1-y^*_2)\in[0,1]$ is the randomization parameter. It is noted that $\phi^*$ in \eqref{eq:optOverall} covers the decision rules in \eqref{eq:LRTex1} (i.e., the LRTs) as special cases for $\nu=0$ or $\nu=1$.

\begin{figure}
\vspace{-0.45cm}
\centering
  \includegraphics[width=0.4\textwidth]{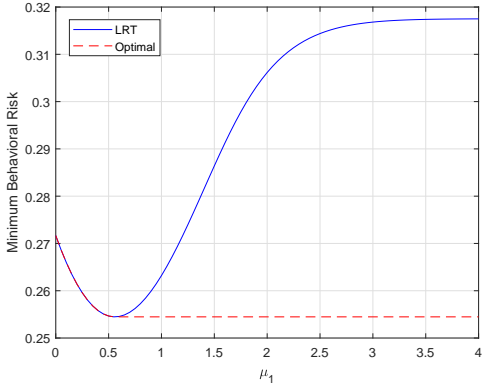}
  \vspace{-0.2cm}
  \caption{{Minimum behavioral risk versus $\mu_1$ for the LRT in \eqref{eq:LRTex1} and the optimal decision rule in \eqref{eq:optOverall} in the Gaussian location testing example.}}
  \label{fig:nonLRT}
\vspace{-0.4cm}
\end{figure}

Let $\mu_0=0$, $\sigma=1$, $\alpha=2$ in \eqref{eq:weight}, and $\pi_0=\pi_1=0.5$ (i.e., equal priors). In addition, consider the following perceived costs: $v(c_{00})=0.5$, $v(c_{10})=1.2$, $v(c_{01})=1$, and $v(c_{11})=0.8$. Then, according to {Proposition 1-Case~(b)}, the LRT may or may not be an optimal solution in this scenario. To observe this fact, consider the minimization problem of the behavioral risk over the LRTs in \eqref{eq:LRTex1} and denote the corresponding minimum behavioral risk as $f^*_{\rm{LRT}}$ (i.e., the solution of \eqref{eq:optProb} over the decision rules in \eqref{eq:LRTex1}). Similarly, let $f^*_{\rm{opt}}$ represent the minimum behavioral risk achieved by \eqref{eq:optOverall}, which actually corresponds to the global solution of \eqref{eq:optProb} due to {Proposition 2}. In Fig.~\ref{fig:nonLRT}, $f^*_{\rm{LRT}}$ and $f^*_{\rm{opt}}$ are plotted versus $\mu_1$.\footnote{In the considered example, $f^*_{\rm{LRT}}$ corresponds to the minimum behavioral risk achieved by the first rule in \eqref{eq:LRTex1} since the second rule yields higher minimum behavioral risks for all values of $\mu_1$.}  The figure reveals that the LRT is not an optimal solution in this example for large values of $\mu_1$ as the optimal decision rule in \eqref{eq:optOverall} achieves strictly lower behavioral risks in that region. For example, {for $\mu_1=1.5$, the minimum behavioral risks achieved by the LRT and the optimal decision rule are $0.2864$ and $0.2545$, respectively, which are obtained by the following decision rules:
\vspace{-0.2cm}\begin{align}\nonumber
&\phi^*_{\rm{LRT}}(r)=\begin{cases}
0\,,&\hspace{-0.1cm}{\rm{if}}~r<1.164\\
1\,,&\hspace{-0.1cm}{\rm{if}}~r\geq 1.164
\end{cases}
\\\nonumber
&\phi^*(r)=0.627
\begin{cases}
0\,,&\hspace{-0.2cm}{\rm{if}}~r<0.4461\\
1\,,&\hspace{-0.2cm}{\rm{if}}~r\geq 0.4461
\end{cases}
+0.373
\begin{cases}
1\,,&\hspace{-0.2cm}{\rm{if}}~r<-0.4461\\
0\,,&\hspace{-0.2cm}{\rm{if}}~r\geq -0.4461
\end{cases}
\end{align}}

\vspace{-0.25cm}

\noindent Since $\phi^*(r)$ above cannot be expressed in the form of an LRT (cf.~\eqref{eq:LRTex1}), the LRT is not optimal for $\mu_1=1.5$. On the other hand, {for $\mu_1<0.55$,} the LRT becomes an optimal solution, as observed from Fig.~\ref{fig:nonLRT}. Hence, it is concluded that the LRT need not always be an optimal solution to the Gaussian location testing problem in the prospect theory based framework, which is in compliance with {Proposition 1-Case~(b)}.

\vspace{-0.2cm}

\subsection{Example 2: Binary Channel}\label{sec:Ex2}

Suppose bit $0$ or bit $1$ is sent over a channel, which flips bit $i$ with probability $\lambda_i$ for $i\in\{0,1\}$. Therefore, when bit $i$ is sent (i.e., under $\mtH_i$), observation $r$ is equal to $i$ with probability $1-\lambda_i$ and equal to $1-i$ with probability $\lambda_i$, where $i\in\{0,1\}$. For this problem, the likelihood ratio, {$L(r)=p_1(r)/p_0(r)$}, becomes equal to $\lambda_1/(1-\lambda_0)$ for $r=0$ and $(1-\lambda_1)/\lambda_0$ for $r=1$. Then, the LRT compares $L(r)$ against a threshold $\eta$ to make a decision as in \eqref{eq:DecRuleLRT1}.\footnote{The LRT in the form of \eqref{eq:DecRuleLRT2} is also considered; however, it is not discussed in the text for brevity as it is not optimal for the parameter setting employed in the example.} Assuming that $\lambda_0+\lambda_1<1$, the \emph{nonrandomized LRT} (i.e., deterministic LRT) can be expressed as follows depending on the value of $\eta$:
\vspace{-0.2cm}\begin{align}\nonumber
&\textrm{If $\eta<\lambda_1/(1-\lambda_0)$:}~\phi_{\rm{LRT}}^{\rm{det}}(r)=1\,,~r\in\{0,1\}\\\nonumber
&\textrm{If $\eta>(1-\lambda_1)/\lambda_0$:}~\phi_{\rm{LRT}}^{\rm{det}}(r)=0\,,~r\in\{0,1\}\\\nonumber
&\textrm{If $\lambda_1/(1-\lambda_0)\leq\eta\leq(1-\lambda_1)/\lambda_0$:}~
\phi_{\rm{LRT}}^{\rm{det}}(r)=
\begin{cases}
1\,,&\hspace{-0.1cm}r=1\\
0\,,&\hspace{-0.1cm}r=0\\
\end{cases}
\end{align}

\vspace{-0.2cm}

\noindent The possible set of false alarm probability ($x$) and detection probability ($y$) pairs that can be achieved via $\phi_{\rm{LRT}}^{\rm{det}}$ consists of $(x=1,y=1)$, $(x=0,y=0)$, and $(x=\lambda_0,y=1-\lambda_1)$. On the other hand, the \emph{randomized LRT} is obtained as

\begin{align}\nonumber
&\textrm{If $\eta<\lambda_1/(1-\lambda_0)$:}~\phi_{\rm{LRT}}^{\rm{rnd}}(r)=1\,,~r\in\{0,1\}\\\nonumber
&\textrm{If $\eta>(1-\lambda_1)/\lambda_0$:}~\phi_{\rm{LRT}}^{\rm{rnd}}(r)=0\,,~r\in\{0,1\}\\\nonumber
&\textrm{If $\lambda_1/(1-\lambda_0)<\eta<(1-\lambda_1)/\lambda_0$:}~
\phi_{\rm{LRT}}^{\rm{rnd}}(r)=
\begin{cases}
1\,,&\hspace{-0.1cm}r=1\\
0\,,&\hspace{-0.1cm}r=0\\
\end{cases}
\\\nonumber
&\textrm{If $\eta=\lambda_1/(1-\lambda_0)$:}~
\phi_{\rm{LRT}}^{\rm{rnd}}(r)=
\begin{cases}
1\,,&\hspace{-0.1cm}r=1\\
\gamma\,,&\hspace{-0.1cm}r=0\\
\end{cases}
\\
\nonumber
&\textrm{If $\eta=(1-\lambda_1)/\lambda_0$:}~
\phi_{\rm{LRT}}^{\rm{rnd}}(r)=
\begin{cases}
\gamma\,,&\hspace{-0.1cm}r=1\\
0\,,&\hspace{-0.1cm}r=0\\
\end{cases}
\end{align}

\vspace{-0.15cm}

\noindent where $\gamma\in[0,1]$ is the randomization constant. The possible set of false alarm probability and detection probability pairs achieved via $\phi_{\rm{LRT}}^{\rm{rnd}}$ can be characterized by the following function (ROC curve) \cite{Poor}:
\vspace{-0.05cm}\begin{gather}
y=\begin{cases}
\frac{1-\lambda_1}{\lambda_0}\,x\,,&{\rm{if}}~0\leq x\leq \lambda_0\\
(1-\lambda_1)+\frac{\lambda_1}{1-\lambda_0}(x-\lambda_0)\,,&{\rm{if}}~\lambda_0< x\leq 1
\end{cases}\,.
\end{gather}

\begin{figure}
\vspace{-0.48cm}
\centering
  \includegraphics[width=0.4\textwidth]{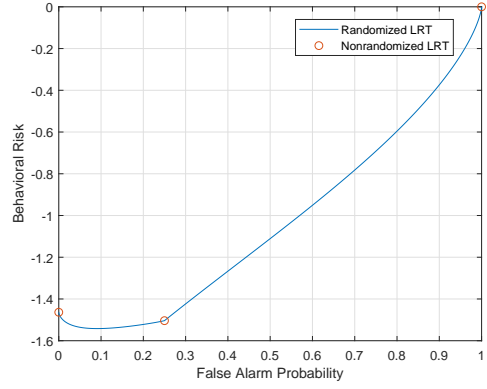}
  \vspace{-0.22cm}
  \caption{{Behavioral risk versus false alarm probability, $x$, for randomized LRT and nonrandomized LRT in the binary channel example.}}
  \label{fig:randLRT}
\vspace{-0.82cm}
\end{figure}

Let $\lambda_0=0.25$, $\lambda_1=0.1$, $\pi_0=\pi_1=0.5$ (i.e., equal priors), {and $\alpha=0.7$ in \eqref{eq:weight}. In addition, consider the following perceived costs: $v(c_{00})=-3$, $v(c_{10})=1.5$, $v(c_{01})=-0.2$, and $v(c_{11})=-1.5$.} Then, based on {Proposition 1-Case~(a),} the LRT is an optimal solution in this scenario. However, in this example, the LRT must employ randomization to achieve the solution of \eqref{eq:optProb}, as stated in {Remark 2}. To illustrate this, Fig.~\ref{fig:randLRT} presents the behavioral risks (see \eqref{eq:BehRisk2}-\eqref{eq:h}) achieved by $\phi_{\rm{LRT}}^{\rm{det}}$ and $\phi_{\rm{LRT}}^{\rm{rnd}}$ with respect to the false alarm probability, $x$. It is observed that the nonrandomized LRT yields the three points marked with circles in the figure, the minimum of which corresponds to a {behavioral risk of $-1.504$.} On the other hand, the randomized LRT achieves the minimum possible {behavioral risk of $-1.542$} (corresponding to the solution of \eqref{eq:optProb}) by employing the following decision rule:
\vspace{-0.1cm}\begin{gather}\label{eq:OptRandSol}
{\phi_{\rm{LRT}}^{\rm{rnd},*}(r)=
\begin{cases}
0.3632\,,&r=1\\
0\,,&r=0\\
\end{cases}}
\end{gather}

\vspace{-0.1cm}

\noindent {The false alarm and detection probabilities of $\phi_{\rm{LRT}}^{\rm{rnd},*}$ are given by $0.0908$ and $0.3269$, respectively,} which are not achievable without randomization. Therefore, it is deduced that the solution of \eqref{eq:optProb} may be in the form of a randomized LRT, which has strictly lower behavioral risk than the optimal nonrandomized LRT, as claimed in {Remark 2}.

{An interesting direction for future work is to specify conditions under which randomization is necessary for LRTs, as mentioned in Remark 2.}


\bibliographystyle{IEEEtran}
\bibliography{HTref}

\end{document}